\documentclass[letter]{aa} 
\usepackage{graphicx}
\usepackage{natbib}
\usepackage{epsfig}
\def\ut#1{\mathop{\vtop{\ialign{##\crcr
     $\hfil\displaystyle{#1}\hfil$\crcr\noalign
     {\kern1pt\nointerlineskip}\hbox{$\hfil\sim\hfil$}\crcr
     \noalign{\kern1pt}}}}}

\def\undersymbol#1#2{\mathop{\vtop{\ialign{##\crcr
     $\hfil\displaystyle{#2}\hfil$\crcr\noalign
     {\kern1pt\nointerlineskip}\hbox{$\hfil#1\hfil$}\crcr
     \noalign{\kern1pt}}}}}

\def\arcmin{^{\prime}}
\def\degr{^0}

\usepackage{graphicx}

\begin{document}

\title{Planck's confirmation of the M31 disk and halo rotation}
       \author{F. De Paolis\inst{1,2}, V.G. Gurzadyan\inst{3}, A.A. Nucita\inst{1,2}, G. Ingrosso\inst{1,2},  A.L. Kashin\inst{3}, H.G. Khachatryan\inst{3}, S. Mirzoyan\inst{3}, E.Poghosian\inst{3}, Ph. Jetzer\inst{4}, A. Qadir\inst{5} \and D. Vetrugno\inst{6}}
              \institute{Dipartimento di Matematica e  Fisica ``E. De Giorgi'', Universit\`a del Salento, Via per Arnesano, I-73100, Lecce, Italy  
              \and INFN, Sez. di Lecce, Via per Arnesano, I-73100, Lecce, Italy
              \and Yerevan Physics Institute and Yerevan State
University, Yerevan, Armenia
\and
Physik-Institut, Universit\"at
Z\"urich, Winterthurerstrasse 190, 8057 Z\"urich, Switzerland
\and
Centre for Advanced Mathematics and Physics,
National University of Sciences and Technology, Rawalpindi,
Pakistan
\and 
Department of Physics, University of Trento, I-38123 Povo, Trento, Italy and 
TIFPA/INFN, I-38123 Povo,  Italy
}

   \offprints{F. De Paolis, \email{francesco.depaolis@le.infn.it}}
   \date{Submitted: XXX; Accepted: XXX}

 \abstract{{\it Planck}'s data acquired during the first 15.4 months of observations  towards both the disk and halo of the M31 galaxy are analyzed. We confirm the existence of a temperature asymmetry, previously detected by using the 7-year WMAP  data,  along the direction of the M31 rotation, therefore indicative of a 
Doppler-induced effect. The asymmetry extends up to about $10\degr$ ($\simeq 130$ kpc) from the M31 center. We also investigate the recent issue raised in Rubin and Loeb (2014) about the
kinetic  
Sunyaev-Zeldovich effect from the diffuse hot gas in the Local Group, predicted to generate a hot spot of a few degrees size in the CMB maps in the direction of M31, where the free electron optical depth gets the maximum value. We also consider the issue whether in the opposite direction with respect to the M31 galaxy the same effect induces a minimum in temperature in the Planck's maps of the sky.  We find that the {\it Planck}'s data at 100 GHz show an effect even larger than that expected.}


   \authorrunning{De Paolis et al.}
   \titlerunning{Possible detection of the M31 rotation in WMAP data}
   \maketitle
%

\section{Introduction}
Galactic disk rotation can be accurately investigated in the optical, infrared (IR) and radio bands
and allows to infer important information, among others, about the dynamical mass content of galaxies (see e.g.
\citealt{binney}). On the other hand, 
many ambiguities still exist about the main constituents of the galactic halos.
The degree to which galactic halos rotate with respect
to the disks is a particularly difficult task to be investigated, even for the closest galaxy to the 
Milky Way: M31 \citep{courteau}.
A novel approach in the study of the rotation of either the disk and halo of nearby galaxies (particularly the M31 galaxy) has been discussed in \cite{depaolis2011}. By using the 7-year WMAP  data, a possible temperature asymmetry was found
both in the M31 disk and halo along the direction of the M31 rotation, therefore reminiscent of a 
Doppler-induced effect.  By adopting the geometry described in Fig. 1 in  \cite{depaolis2011}, and extending the analysis up
to about $20 \degr$ ($\simeq 260$ kpc) around the M31 center, we found in the two opposite regions of the M31 disk a temperature difference of about $130$ $\mu$K, more or less the same in the  W, V and Q WMAP bands.
A similar effect was visible also towards the M31 halo up to about 120 kpc from the M31 center with a peak value of about $40$ $\mu$K. The robustness of that result was tested 
by considering 500 randomly distributed control fields and also simulating 500 sky map realizations from the best-fit power spectrum constrained with BAO and $H_0$ (see \citealt{depaolis2011} for details).
It turned out that the probability that the detected temperature asymmetry towards the M31 disk is due to a random fluctuation of the CMB signal is below about $2\%$ while in the case of the M31 halo
it is less than about $30\%$. Although the confidence level of the signal was not high with WMAP data, if estimated purely statistics, nevertheless we believed that the geometrical structure of the temperature asymmetry
pointed towards a definite effect modulated by the rotation of the M31 disk and halo and suggested that with the {\it Planck} data it could be possible to definitely prove or disprove our conclusions. Indeed, the {\it Planck} satellite is about ten times
more sensitive than the WMAP satellite and has an angular resolution about three times better:
the {\it Planck} full width half maximum (FWHM) resolution ranges from $33.3 \arcmin$ to $4.3 \arcmin$ going from 30 GHz to 857 GHz, and its final sensitivity is in the range of  $2-4.7$ $\mu$K/K in terms of 
$\delta T/T$ for the Low Frequency Instrument (LFI), that is in the range $30-70$ GHz, and of $2-14$  $\mu$K/K  for the High Frequency Instrument  (HFI) below 353 GHz 
(see e.g.  \citealt{burigana2013} for a recent review on {\it Planck}'s results). 
The  aim of the present paper is therefore to analyze in detail the {\it Planck} data acquired during the first 15.4 months of observations  towards both the disk and halo of the M31 galaxy.
In addition, we also take the opportunity of investigating in some detail the  recent issue raised in \cite{rubinloeb} about  the kinetic Sunyaev-Zel'dovich effect from the diffuse hot gas in the Local Group, which happens to show up as a hot spot of a few degrees in size in the direction of M31 (where the free electron optical depth gets the maximum value). We also investigate whether in the opposite direction with respect to the M31 galaxy, the same effect induces a minimum in temperature in the {\it Planck}'s maps of the sky.

\section{Planck analysis}
Two instruments are present onboard the {\it Planck} satellite: the LFI \citep{Bersanellietal2010} covers the 30, 44, and 70 GHz bands by using amplifiers cooled to 20 K. 
The HFI \citep{Lamarreetal2010} covers the 100, 143, 217, 353, 545, and 857 GHz bands with bolometers cooled down to 0.1 K. 
{\it Planck}'s sensitivity, angular resolution (from $3 \arcmin$ to $5 \arcmin$)
and frequency coverage make it a powerful instrument for cosmology and both galactic and extragalactic astrophysics \citep{planckI}.
In order to reveal the different contribution by the M31 disk and
halo, the region of the sky around the M31 galaxy has been divided
into several concentric circular areas as shown in Fig. 1 in 
\cite{depaolis2011}, to which we refer for further details. Here we only mention that the M1 region is the M31 south-east half-disk while the M2 region corresponds to the north-west half-disk. Since the M31 disk is rotating  in the clock-wise direction we expect that the M1 region would be hotter than the M2 one. 
The mean
temperature excess $T_m$ in $\mu$K in each
region has been obtained in each {\it Planck}'s band and is shown in Table 1 with
the corresponding standard error (SE)\footnote{The standard error given in the fourth column is
calculated as the standard deviation of the excess temperature
distribution divided by the square root of the pixel number in
each region. To possibly enable the comparison with the previous WMAP data analysis \cite{depaolis2011}, here we use {\it Planck}'s 100 GHz data 
and have verified that, within the errors, the sigma
values calculated in that way are consistent with those evaluated
by using the covariance matrix obtained by a best fitting
procedure with a Gaussian to the same distribution. In the last column we give the average excess temperature for 360 control fields with the usual standard deviation.}, along with
the number of pixels in each area.
\begin{table*}[tbp]
\begin{center}
{
\renewcommand{\baselinestretch}{1.2}
\renewcommand{\tabcolsep}{2.6mm}
\small
\centering
\begin{tabular}{ccrccc}
\hline \hline
 R, deg, kpc & Region   & N, pix &   $T_m\pm SE$  & $T_m\pm \sigma$ for 360 control fields \\
\hline
  1.5,  19.5 & M1       &   4213 & $115.4\pm 2.0$ & $44.0\pm 5.0$   \\
  1.5,  19.5 & M2       &   4182 & $48.2\pm 2.0$  & $50.0\pm 6.0 $  \\
  4.0,  51.9 & M1       &  29076 & $70.1\pm 0.9$  & $41.0\pm 7.0$   \\
  4.0,  51.9 & M2       &  28983 & $32.0\pm 0.9$  & $66.0\pm 10.0 $ \\
\hline
  4.0,  51.9 & N1+S1    &  27957 & $70.0\pm 1.0$  & $41.0\pm 7.0$   \\
  4.0,  51.9 & N2+S2    &  27874 & $32.2\pm 1.0$  & $66.0\pm 9.0$   \\
 10.0, 131.2 & N1+S1    & 158752 & $65.0\pm 0.2$  & $43.0\pm 8.0$   \\
 10.0, 131.2 & N2+S2    & 158720 & $52.2\pm 0.2$  & $73.0\pm 10.0$  \\
\hline
  4.0,  51.9 & M31      &  61306 & $50.1\pm 0.3$  & $44.6\pm 1.6$ \\
  4.0,  51.9 & anti M31 &  61306 & $-6.2\pm 0.3$  & $24.8\pm 1.0$  \\
\hline
\end{tabular}
}
\caption{Temperature excess in the M31 regions. The radius of the
considered annulus is given in degrees and in kpc in the first
column; the value of 744 kpc \citep{vilardell} is adopted for the
distance to M31. The second column indicates the considered
region. In the third column the numbers
of pixels in each region are given. The fourth column show
the CMB mean temperature of each region (in $\mu$K)  in
the 100 GHz {\it Planck} map with the corresponding standard error (SE) while the last column gives the average temperature excess in the 360 control fields with the standard deviation (see text for details).}
\end{center}
\end{table*}

\subsection{Results for the M31 disk}
As far as the M31 disk is concerned and as can be seen from the first four lines of Table 1 and Fig. 1,
the M1 region turns out to be  always hotter than the M2 region. For example, at $1.5\degr$ the M1 region is 
67 $\mu$K hotter than the M2 region and even at $4\degr$ the M1 region is 38 $\mu$K hotter than M2.  The upper panel of Fig. 1 clearly shows the temperature asymmetry profile in the two regions of the M31 disk. These profiles are in agreement with the  results obtained previously by using the WMAP data \citep{depaolis2011}. 
\footnote{The absence of foreground reduced  Planck's maps, that where instead available for WMAP maps, makes ambiguous the comparison between real and simulated data.
The strategy adopted here of using 360 control fields in the Planck's maps gives more reliable results.}
Moreover, the shape of the two profiles is clearly mirror-like, as expected if the effect is due to a Doppler modulation induced by the M31 disk rotation. Indeed, the hotter (M1) region
corresponds to the side of the M31 disk that rotates towards us. This mirror-like shape of the two regions of the M31 disk is also visible in the  M31 thick HI  disk obtained at 21 cm \citep{Chemin,Corbelli}.
In order to test whether the temperature asymmetry we see towards the M31 disk is real or can be explained as a random fluctuation of the CMB signal (that indeed is rather patchy) we adopt a different strategy with respect to that in \cite{depaolis2011}. We consider 360 control field regions with the same shape as the M1 and M2 regions and at the same latitude as M31 but at  $1\degr$ longitude from each other. For each region we determine the excess temperature profile and calculate the average profile and the corresponding standard deviation. As can be easily observed by looking at  Table 1 and at the bottom panel of Fig. 1, the M1 region for the 360 control fields is always cooler than the M2 region, exactly the opposite of what happens towards the M31 disk. Moreover, the M1 temperature towards M31  is always significantly larger with respect to the corresponding temperature of the control fields and even at 4 degrees the effect is at $\simeq 4\sigma$. The same also holds for the M2 region: the 360 control fields have a temperature excess of  $66\pm 10$ $\mu$K at $4\degr$  while the temperature of the M31 M2 region is always cooler, being $\simeq 32$ $\mu$K. The effect is therefore at $\simeq 3\sigma$ at $4\degr$.
We remark that we have conducted the same study in all the {\it Planck} bands and find that the results, presented for convenience only for the 100 GHz band in this paper, are comparable in each band.

\subsection{Results for the M31 halo}

Adopting the same geometry as in \cite{depaolis2011}, we have estimated the temperature excess in the {\it Planck}'s sky maps in the N1+S1 region (the region in the south-east of the M31 halo that is expected to be rotating moving towards the Milky way, if the M31 halo is rotating along the same axis of the disk) and in the N2+S2 region (the opposite region with respect to the rotation axis). As can be seen from Table 1 and Fig. 2 (upper panel), the N1+S1 region turns out to be hotter than the N2+S2 one at any galactocentric distance. The temperature difference peaks at about $4\degr$ (with a value about 38 $\mu$K), but continues up to $10\degr$ (where it is still at $\simeq 13$ $\mu$K). Beyond about $12\degr$ the temperature asymmetry gets inverted and the N2+S2 region becomes hotter than the N1+S1 one, as a result of the intersection with the Milky Way disk that clearly shows up in the CMB maps. As for the M31 disk, also for the  halo one can observe a kind of mirror symmetry between the  N1+S1 and  N2+S2 regions, although less pronounced with respect to the case of the M31 disk. We have also tested whether the measured temperature asymmetry is due to a random fluctuation of the CMB signal by considering 360 control fields with the same shape of the  N1+S1 and  N2+S2 regions at the same latitude of M31 but at different longitude values (the control  fields are equally spaced at one degree distance each other in longitude). As one can see from the bottom panel of Fig. 2, also for the halo regions (as for the M31 disk) the temperature asymmetry in the 360 control fields clearly has an opposite behavior with respect to the profiles towards  M31 and the N1+S1 regions are always cooler than the N2+S2 ones (bottom panel in Fig. 2). This effect is clearly due to the presence of the Milky Way disk in the CMB sky maps which makes the  N2+S2 regions generally hotter than the N1+S1 regions.
It can be easily observed by comparing the temperature asymmetry profile of the M31 halo with that of the control fields that the N1+S1 region of M31 is always hotter than the control fields profile (with a confidence level of about   $4\sigma$ at $4\degr$ and $2.7\sigma$ at $10\degr$) while the N1+S1 region  is cooler than the control fields profile (with a confidence level of about   $3.7\sigma$ at $4\degr$ and $2.1\sigma$ at $10\degr$). We can therefore conclude that the probability  that the asymmetry effect towards the M31 halo at 100 GHz is due to a random fluctuation of the CMB signal is well below $1\%$.
We also point out that we have verified that the temperature
asymmetry towards the M31 halo vanishes if the adopted geometry is
rigidly rotated by an angle larger than about $10\degr$ with
respect to the assumed M31 rotation  axis, thus giving a further indication
that the asymmetric halo temperature is a  genuine effect due to the halo rotation and not
simply a random fluctuation of the CMB signal.
\begin{figure}[h!]
 \centering
  \includegraphics[width=0.44\textwidth]{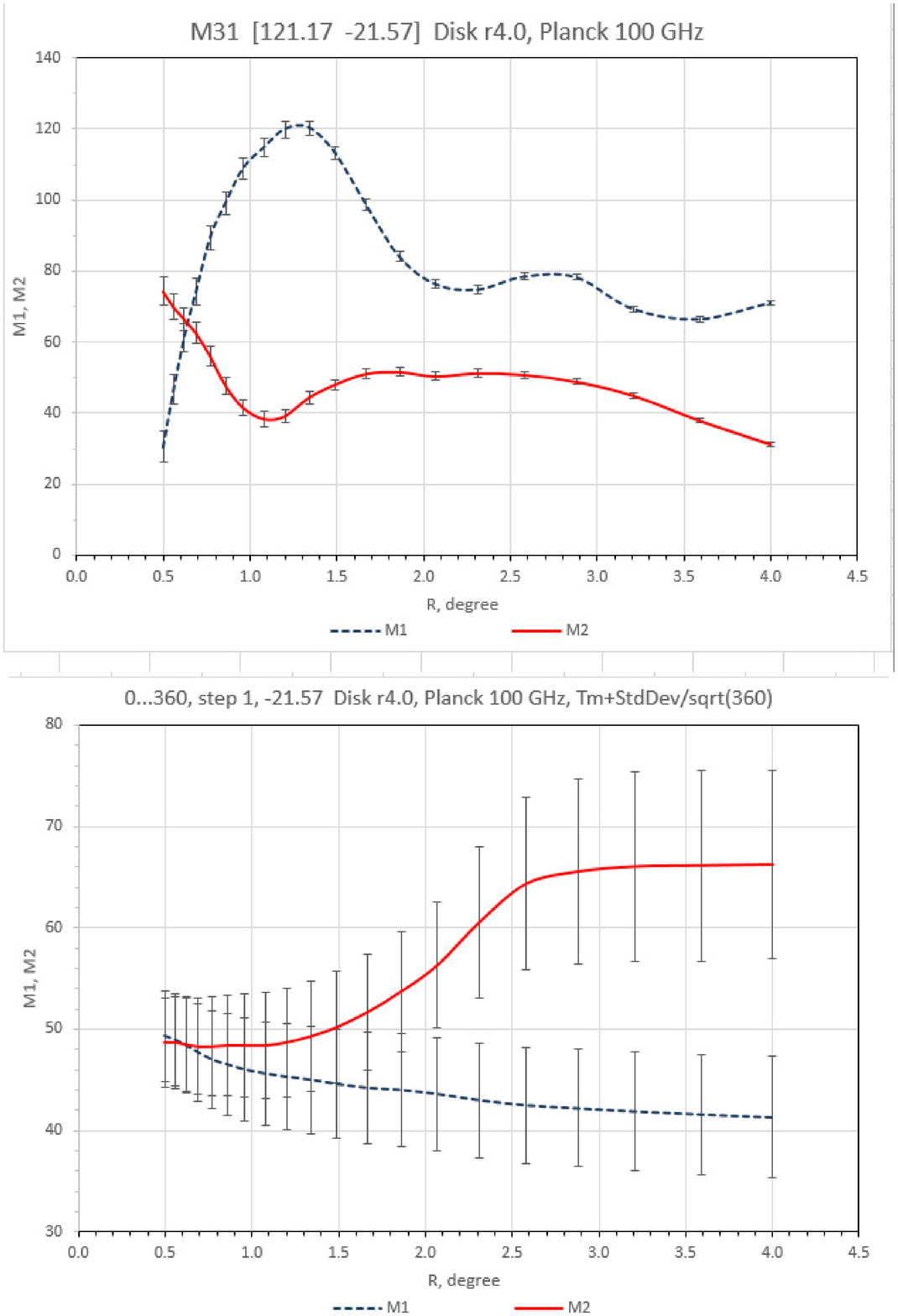}
 \caption{Upper panel: the excess temperature profiles (in $\mu$K) for the M1 and M2 regions of the M31 disk. Bottom panel: temperature profiles (in $\mu$K) for 360 regions equally spaced at one degree distance each other in longitude and at the same latitude as M31.}
 \end{figure}
\begin{figure}[h!]
 \centering
  \includegraphics[width=0.44\textwidth]{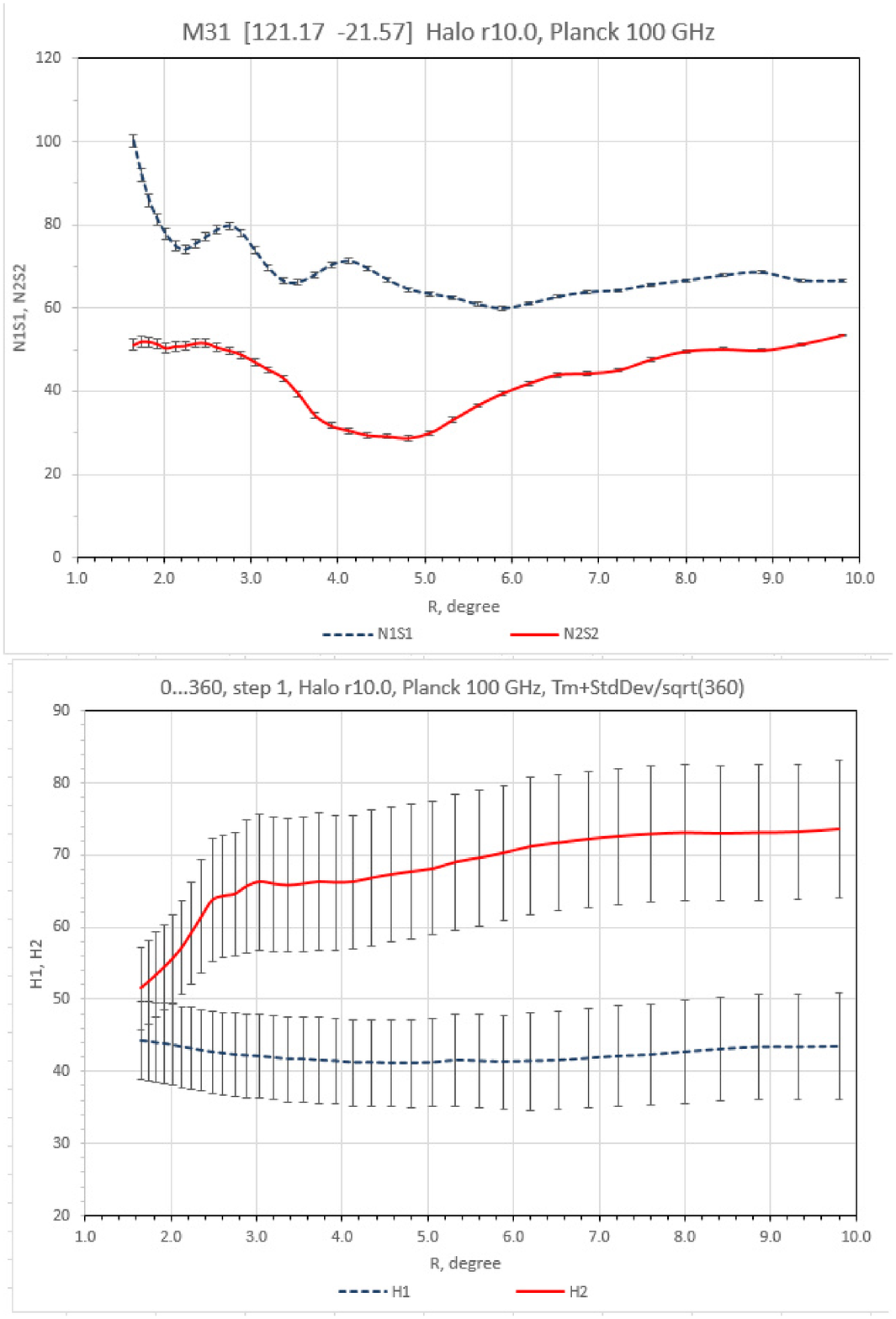}
 \caption{The same as for Fig. 1 but for the M31 halo (upper panel)  and for  360 regions at one degree longitudinal distance each other (H=N+S, bottom panel).}
 \end{figure}
\subsection{The Local Group hot gas effect}
Recently  \cite{rubinloeb} raised an interesting issue related to the kinetic Sunyaev-Zel'dovich effect from the diffuse hot gas in the Local Group.  In fact, since the Local Group moves with respect to the CMB 
\citep{RauzyGurzadyan1998}, its hot gas halo component should imprint a non-primordial temperature shift in the CMB signal. The expected effect should show up as a hot spot of a few degree in size in the direction of the M31 galaxy, which happen to be  opposite with respect to the center of the Local Group. In fact, due to geometrical consideration, the free electron optical depth gets the maximum value just towards the M31 galaxy. On the other hand,  in the opposite direction with respect to the M31 galaxy, the same effect should induce a minimum in temperature in the {\it Planck}'s maps of the sky.  We have investigated this issue by looking at the  
{\it Planck}'s sky map at 100 GHz and find (see the last two lines at the bottom of Table 1) that the mean temperature excess in a $4\degr$ circle towards the M31 galaxy is $\simeq 50.1\pm 0.3$ $\mu$K, therefore consistently hotter with respect to the average temperature in the southern hemisphere of the sky ($\simeq 40.5$ $\mu$K) \footnote{The mean temperature in both the southern and northern hemispheres of the sky has been evaluated after the equatorial region, affected by the Milky Way emission, has been subtracted}. Towards the anti M31 direction ($l=301.17\degr$, $b=21.57\degr$)  we find,  in a $4\degr$ circle, a temperature excess of  $-6.2\pm 0.3$ $\mu$K, to be compared to the average  temperature in the northern hemisphere of the sky of about  $27.5$ $\mu$K. 
As for the M31 disk and halo, we also consider 360 control fields -- this time randomly extracted around (within $20\degr$) either the M31 and anti M31 directions.  The obtained results are shown in the two lines at the bottom of Table 1 and, as one can see, the $4\degr$ circle towards M31 is hotter than the control fields at about $3\sigma$. Towards the anti M31 direction, the $4\degr$ circle is clearly much cooler with respect to the control fields (at about $29\sigma$).
The effect predicted towards the M31 galaxy by \cite{rubinloeb} was of a few $\mu$K and no mentioning   about the possible existence of a cold spot in the anti M31 direction was present there.  From the discussion above it is clear that  the observed temperature difference (about $56$ $\mu$K) between the M31 and anti M31 directions cannot be explained as a random fluctuation of the CMB signal and should therefore arise from two main contributions: the kinetic Sunyaev-Zel'dovich effect  and the presence of hot gas in the M31 halo, with  a density larger than the average hot gas density in the Local Group. The presence of this hot gas halo component in the M31 halo, as predicted  in \cite{paperdijqr95}, might  be able to explain both the CMB temperature increase towards the M31 galaxy and, if it rotates around the same rotation axis as the M31 disk, the temperature shifts between the two sides of the M31 halo as discussed above.

\section{Conclusions}

Galactic halos
are relatively less studied than  galactic disks and there are still many
ambiguities not only in the main halo constituents, but also  with respect to the degree to which galactic halos rotates with respect to the
disks \citep{courteau,deason}.   Actually, 
the rotation of the galactic halos is clearly related to the
formation scenario of galaxies. In the standard collapse model
(see e.g. \citealt{eggen}) both the halo and disk  derive
from the same population and the rotation  of the outer halo should
be, in this case, aligned with the disk angular momentum. On the
contrary, in a hierarchical formation scenario, structures
reaching later the outer halo should be less connected
to the disk. Therefore, it is evident that information on the
galactic halo rotation provides key insights about the formation
history  of galaxies. 
It is also well known that the M31 disk rotates
with a speed of about 250 km s$^{-1}$ and this has been clearly
shown also by the velocity maps obtained from radio measurements 
\citep{Chemin,Corbelli}.
These maps look  very similar to what we find in the {\it Planck} data towards the M31 disk.
In the previous Section we have also shown that {\it Planck}'s data show the existence of a  temperature asymmetry with respect to the disk-halo rotation axis, up to a galactocentric distance of about 130 kpc and  with a peak  temperature contrast of about $40~\mu$K. We remark that, until now, the only evidence of the M31 halo rotation was put forward by the analysis of the dwarf galaxies orbiting M31 \citep{ibata}.

In all generality, five possibilities may be considered in order to explain the effects discussed in Sections 2.1-2.3: ($i$)
free-free emission; ($ii$) synchrotron emission; ($iii$) anomalous microwave emission (AME) from dust grains;  ($iv$) kinetic
Sunyaev-Zel'dovich (SZ) effect; ($v$) cold gas clouds
populating the M31 halo. A detailed study of their contribution using all the {\it Planck}'s bands to constrain the model parameters and the relative weight of these five models will be published elsewhere. Here, we only note that effects $(i)-(iii)$ give a signal with a rather strong dependence on the wavelength, while $(iv)$ and $(v)$ are almost independent of the observation band in the microwave regime and to first approximation could provide the main contribution to the observed effect.
Thus, our investigation shows the power of CMB to trace, along with the clusters of galaxies via Sunyaev-Zeldovich effect and the large scale voids (e.g. \citealt{G}), also the individual galactic halos. 

\begin{acknowledgements}
{We acknowledge the use of the Legacy Archive for
Microwave Background Data Analysis (LAMBDA) and HEALPix
\citep{gorski/etal:2005} package. PJ acknowledges support from the
Swiss National Science Foundation.}
\end{acknowledgements}

\end{document}